# A New Scheduling Algorithms
# for Real Time Tasks

C.Yaashuwanth

Department of Electrical and Electronics Engineering,
Anna University Chennai, Chennai 600 025, India.
.

Dr.R.Ramesh

Department of Electrical and Electronics Engineering,
Anna University Chennai, Chennai 600 025, India.
.

*Abstract -* **The main objective of this paper is to develop the two different ways in which round robin architecture is modified and made suitable to be implemented in real time and embedded systems. The scheduling algorithms plays a significant role in the design of real time embedded systems. Simple round robin architecture is not efficient to be implemented in embedded systems because of higher context switch rate, larger waiting time and larger response time. Missing of deadlines will degrade the system performance in soft real time systems. The main objectives of this paper is to develop the scheduling algorithms which removes the drawbacks in simple round robin architecture. A comparison with round robin architecture to the proposed architectures has been made. It is observed that the proposed architectures solves the problems encountered in round robin architecture in soft real time by decreasing the number of context switches waiting time and response time thereby increasing the system throughput.**

*Key words - Round robin architecture, time slice, deadlines, soft real time.*

## I. INTRODUCTION

This paper describes the two different ways in which round robin architecture is modified and made suitable to be implemented in real time and embedded systems

The new approach describes the scheduling of processes in modified round robin architecture. Simple round robin architecture is not efficient if the taskset has processes with variable CPU burst because the processes arriving in round robin will be allocated time slice in first come first serve manner. This drawbacks of variable process bursts of processes in a taskset leads to larger waiting time. The larger response time of processes with less CPU burst which leads to increase in waiting time and response time of system thereby degrading the system performance The proposed algorithm has the advantage of processes with less CPU burst will have a better waiting time and response time than compared to simple round robin architecture and covers the drawbacks of round robin architecture by reducing the waiting time and reducing the response time thereby increasing the system throughput

The second approach describes the calculation of intelligent time slice for round robin architecture is a modified version of simple round robin architecture. Simple round robin architecture cannot be implemented in soft real time systems because of high context switch rate, larger waiting time and larger response time. Because of these performance criteria of the round robin architecture is not suitable to implement in real time systems. Soft real time systems have no hard deadlines for tasks but missing of deadlines in soft real time systems will degrade the system performance.

Real time systems always has a time constraint on computation. Each task should be invoked after the ready time and must complete before its deadline[12][13][14], an attempt has been made to satisfy these constraints. Simple round robin architecture[1] is not suitable to implement in softreal time due to more no of context switches, longer waiting and response times. This in turn leads to low throughput in the system . Richard roehi[3] proposed a new way of scheduling which implements a new priority queue in the round robin architecture that gives priority to tasks with short CPU burst thereby improving the performance of the tasks with less central processing unit (CPU) burst. Fair scheduling with tunable latency[8] is a Round Robin approach that proposes an alternative and lower complexity approach to packet scheduling, based on modifications of the classical Round Robin scheduler. The authors showed that appropriate modifications of the weighted Round Robin (WRR) service discipline can, in fact, provide tight fairness properties and efficient delay guarantees to multiple sessions. Round robin approach has its applications on networks by allowing the network devices to have a free share of network resources. Various types of scheduling algorithms such as Dificit round robin [6], Dificit round robin alternated [4] and credit round robin[7] have been implemented.

Real-time scheduling theory has shown a transition from cyclical executive based infrastructure to a more flexible scheduling models such as fixed-priority scheduling, dynamic-priority scheduling, feedback scheduling or extended scheduling[10]. In fact, recent studies show that almost every existing real-time operating system provides only POSIX-compliant fixed-priority scheduling [11] since it can be easily implemented in commercial kernels,. In task scheduling policies for real time system the author[1] has discussed the basic architecture of fixed priority scheduling which implies that there will be single server and the jobs will be allocated based on the priority given by the user. A. Burns[5] proved that standard analysis of fixed priority assumes that all the computations of each task must be completed in task deadlines but in practice this is not the case, deadlines is most currently associated with last observed event of the task. Reindir and Pieter revealed[2] the worst case response time analysis of real time tasks using hierchial fixed priority scheduling .Using an example which consist of single server the author portrays that the existing worst case response time analysis can be improved.

With these observations it is found that these existing simple round robin architecture are not suitable to be implemented realtime embedded system . The proposed algorithms are a modified version of round robin algorithm with shortest jobs scheduling of tasks in round robin and intelligent time slicing concept .

## II.  ROUND ROBIN ARCHITECTURE

Round robin architecture is a pre emptive version of first come first serve scheduling algorithm. The processes are arranged in the ready queue in first come first serve manner and the processor executes the process from the ready queue based on time slice. If the time slice ends and the processes are still executing on the processor the scheduler will forcibly pre-empt the executing process and keeps it at the end of ready queue then the scheduler will allocate the processor to the next process in the ready queue. The preempted process will make its way to the beginning of the ready list and will be executed by the processor from the point of interruption.

A scheduler requires a time management function to implement the round robin architecture and requires a tick timer. The time slice is proportional to the period of clock ticks. The time slice length is critical issue in soft real time embedded application as missing of deadlines will have negligible effects in the system performance. The time slice must not be too small which results in frequent context switches and should be slightly greater than average process computation time.

### A.  Round Robin Drawbacks in Operating Systems

Round robin when implemented in soft real time systems faces two drawbacks they are high rate of context switch and low throughput. These two problems of round robin architecture are interrelated.

### B.  Larger waiting time and Response time

In round robin architecture the time the process spends in the ready queue waiting for the processor to get executed is known as waiting time and the time the process completes its job and exits from the taskset is called as turn around time. Larger waiting and response times are clearly a drawback in round robin architecture as it leads to degradation of system performance.

### C.  Low throughput

Throughput is defined as number of process completed per time unit. If round robin is implemented in soft real time systems throughput will be low which leads to severe degradation of system performance. If the number of context switches is low then the throughput will be high. Context switch and throughput are inversely proportional to each other.

### D.  Context Switch

When the time slice of the task ends and the task is still executing on the processor the scheduler forcibly pre empts the tasks on the processor  and stores the task context in stack or registers and allocates the processor to the next task in the ready queue. This action which is performed by the scheduler is called as context switch. Context switch leads to the wastage of time, memory and leads to  scheduler overhead.

These drawbacks are eliminated in the modified version of round robin described in the next sections.

## III.  MODIFIED ROUND ROBIN ALGORITHMS FOR REAL TIME EMBEDDED SYSTEM

### A.  Shortest round Robin Architecture

The proposed architecture focuses on the drawbacks of simple round robin architecture which gives equal priority to all the processes (processes are scheduled in first come first serve manner) Because of this drawback round robin architecture is not efficient for processes with smaller CPU burst. This results in the increase in waiting time and response time of processes which results in the decrease in the system throughput.

The proposed architecture eliminates the defects of implementing a simple round robin architecture in by scheduling of processes based on the  CPU burst, A dedicated small processor used to reduce the burden of the main processor is assigned for rearranging of processes in the ascending order based on the CPU burst of the process (lower to higher) The proposed architecture has greater waiting time response time and throughput thereby improving the system performance. The process scheduling in proposed architecture is shown in figure 3.1

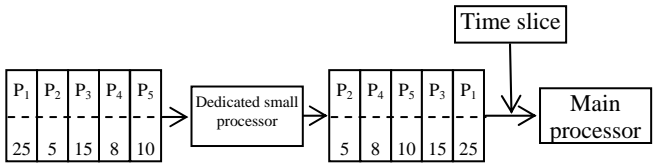

Figure 3.1 Process scheduling in shortest round robin

### B. *Intelligent time slice for Round robin in soft real time scheduling*

The proposed architecture focuses on the drawbacks of simple round robin architecture which are context switch, equal priority to all the tasks Because of these drawbacks round robin architecture is not suitable for softreal time systems. In soft real time systems the missing of task deadlines are not catastrophic but have a negligible effects on the system which results in degradation of system performance.

The proposed architecture eliminates the defects of implementing a simple round robin architecture in softreal time system by introducing a concept called intelligent time slicing which depends on three aspects they are priority, average CPU burst, context switch avoidance time. The proposed architecture allows the user is allowed to assign priority to the system . An assumption is made on average CPU burst which are reasonable to the system. A dedicated small processor used to reduce the burden of the main processor is assigned for calculating the time slice. The calculated time slice will be different and independent for each tasks and the tasks are fed into the ready queue and these tasks execute in the main processor with their individual time slices. The proposed architecture can be implemented in softreal time because of greater response time and throughput and the users can allocate priority to every individual task. The intelligent time slice are calculated by the dedicated small processor as shown in figure.4.1

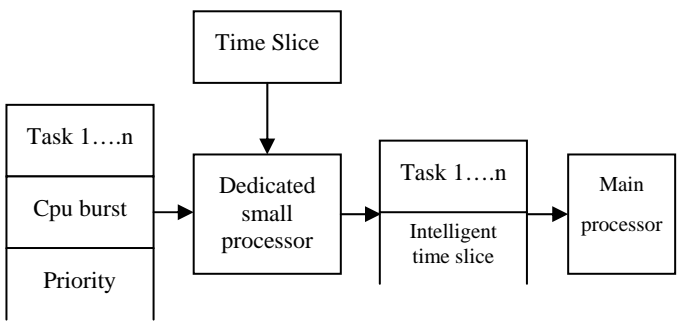

Figure 3.2 Intelligent time slice generation.

The priority, average CPU burst and context switch avoidance time along with original time slice is given to the small dedicated processor which calculates the new time slice dedicated to the corresponding tasks. The output of small dedicated processor will be the task id no, CPU burst and the calculated time slice these criteria are given to the main processor. These dedicated time slice are different and are exclusively allocated for each tasks and the tasks execute on the processor based on these time slicing. Thus the proposed architecture is superior to existing simple round robin architecture and can be implemented in soft real time systems.

### *Intelligent Time Slice Calculation*

A new way of intelligent time slice calculation has been proposed which allocates the frame exclusively for each task based on priority, shortest CPU burst time and context switch avoidance time.

Let the original time slice (OTS) is the time slice to be given to any process if it deserves no special consideration.

> Intelligent time slice = Original Time Slice(OTS)+ Priority Component (PC)+ Shortness Component for CPU burst time (SC)+ Context Switch Component(CSC)

----------→equation 1

Priority component (PC) is assigned by the user depending upon the priority which is inversely proportional to the priority number (higher the priority, greater the PC).

Shortness component (SC) is assigned inversely proportional to the length of the next CPU burst for the process. Shortest component should be lesser than assumed CPU burst (ATS).

Context Switch Component(CSC) is calculated as follows:

o Calculate the Computed Component (CC): Priority Component (PC) and Shortness Component (SC) are added to the Original Time Slice (OTS).

> Computed component (CC) = Priority Component(PC) + Shortness Component (SC) + Original Time Slice (OTS)

o CC is deducted from the Assumed CPU burst (ATS). Let the result be called balance CPU burst.

> Balanced CPU burst = Assumed Time Slice(ATS) − Computed Component (CC)

If this balance CPU burst is less than OTS, it will be considered as Context Switch Component (CSC).

### IV. CASE STUDY

Five processes has been defined (with its priority), these five processes are scheduled in round robin and also in the proposed architecture. The context switch, waiting time, turn around time has been calculated and the results were compared. The process id, burst time and priority are defined as shown in table 4.1

TABLE 4.1 INPUT COMPONENT FOR THE PROCESSOR

| Process ID | CPU burst time (milliseconds) | Priority |
|---|---|---|
| 1 | 25 | 2 |
| 2 | 5 | 3 |
| 3 | 15 | 1 |
| 4 | 8 | 2 |
| 5 | 10 | 1 |

Time slice = 4m.sec.

A. *Round robin architecture*

The above five processes has been scheduled using simple Round Robin architecture. The time slice of four milliseconds has been used. In round robin algorithm no process is allocated CPU for more than one time slice in a row. If the CPU process exceeds one time slice, the concern process will be preempted and put into the ready queue. The process is preempted after the first time quantum and the CPU is given to the next process which is in the ready queue (process P2), similarly schedules all the process and completes the first cycle In the second cycle process P2 requires one millisecond time slice (doesn't require four milliseconds time slice), so it quits before its time quantum expires. The CPU is given to next process P3. In the same way all other processes in the system are scheduled once each process have received one time slice the CPU is returned to process P1 for additional time slice. The process time slicing in simple Round Robin architecture is shown in figure.5.1

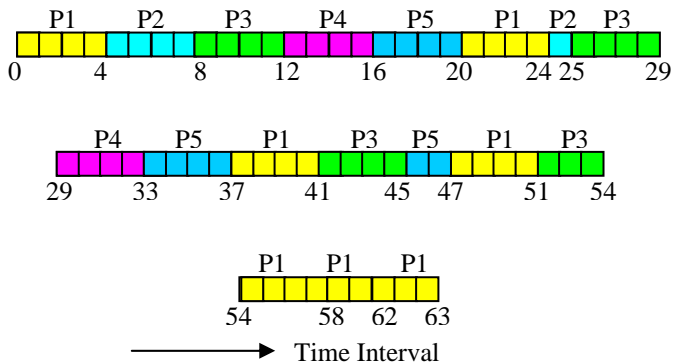

Figure 4.1 Time slicing in Round Robin architecture

B. *Shortest round robin architecture*

The figure 4.2 shows the scheduling of five processes using the proposed architecture

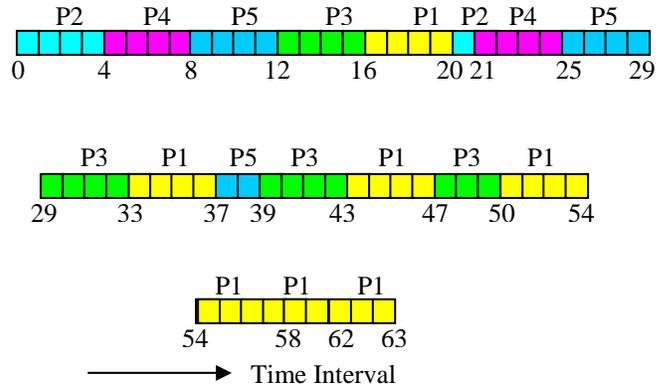

Figure 4.2 Proposed architecture

The above five process has been scheduled by the proposed architecture with time slice of four milliseconds . process P2 is allocated first to the processor since it has the smallest CPU burst. It executes its time slice of 4 milliseconds and is pre empted by the processor. The processor is then allocated to next process with smallest CPU burst process P4. in the same way processor completes the cycle by allocating one unit time slice to all the process. In the second cycle the process p2 has only one millisecond to execute, after completing the execution of process p2 the processor is pre empted and allocated to process P4 and the cycle is repeated in round robin fashion till all the tasks complete their CPU burst.

C. *Intelligent time slice for round robin architecture*

The above five processes has been scheduled using proposed architecture. Intelligent time slice has been calculated using the equation-1 and the calculated intelligent time slice is shown in table 4.2

TABLE 4.2 CALCULATION OF INTELLIGENT TIME SLICE FOR ROUND ROBIN ARCHITECTURE

| Process ID | CPU burst | Calculated time slice (milliseconds) | | | | |
|---|---|---|---|---|---|---|
| | | OTS | PC | SC | CSC | Intelligent time slice |
| 1 | 25 | 4 | 0 | 0 | 0 | 4 |
| 2 | 5 | 4 | 0 | 1 | 0 | 5 |
| 3 | 15 | 4 | 1 | 0 | 0 | 5 |
| 4 | 8 | 4 | 0 | 1 | 3 | 8 |
| 5 | 10 | 4 | 1 | 0 | 0 | 5 |

The intelligent time slice of process P1 is same as the original time slice of four milliseconds and time slice of four milliseconds is assigned to process P1. After the execution of four milliseconds time slice the CPU is allocated to process P2. Sine the CPU burst of process P2 is lesser than the assumed CPU burst (ATS), one milliseconds of SC has been included. The process P3 has the highest priority, so priority component is added and the total of five milliseconds is allocated to process P3. The Balanced CPU burst for process P4 is leaser than OTS, context switch component is added and a total of eight millisecond time slice is given to process P4. Process P5 is given a total of five milliseconds with one millisecond of priority component is added to original time slice. After executing a cycle the processor will again be allocated to process P1 for the next cycle and continuously schedules in the same manner. The process time slicing in proposed architecture is shown in figure 4.3

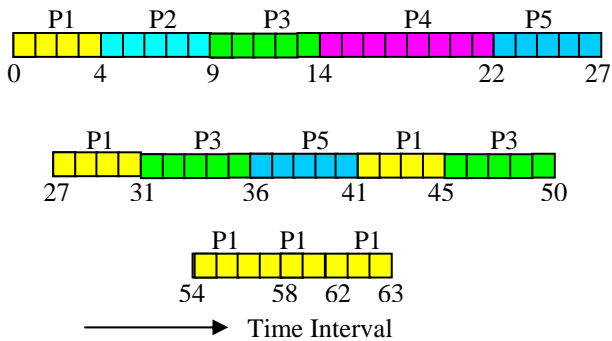

Figure 4.3 Time slicing in Proposed architecture 2

Waiting time and turn around time has been calculated using the formula given below and the results were compared.

Waiting time is calculated by

$$\sum_{i=1}^{n} \frac{wt_i}{n}$$

wt $\rightarrow$ waiting time of process.
n $\rightarrow$ No. of process.

Turn around time is calculated by

$$\sum_{i=1}^{n} \frac{tt_i}{n}$$

tt $\rightarrow$ turned around time of process.
n $\rightarrow$ No. of process.

The computed results are tabulated and tabulated in table 4.2

### TABLE 4.3 COMPARISON ROUND ROBIN SHORTEST ROUND ROBIN

| Algorithm | Waiting time in milliseconds | Turn around time in milliseconds |
|---|---|---|
| Simple round robin | 31 | 44 |
| Shortest round robin | 22 | 36 |
| Intelligent time slice for round robin | 25 | 37 |

## V. CONCLUSION AND FUTURE WORKS

A comparative study of round robin architecture shortest round robin and intelligent time slice for round robin architecture is made. It is concluded that the proposed architectures are superior as it has less waiting, response times, usually less preemption and context switching thereby reducing the overhead and saving of memory space. Future work can be based on this architectures modified and implemented for hard real time system where hard deadline systems require partial outputs to prevent catastrophic events.

## About the authors

Mr. C.Yaashuwanth completed his B.Tech. degree in Information technology at BSA CRESCENT Engineering College Chennai, He completed M.E. in Embedded System Technologies at VEL TECH Engineering College Chennai

Dr. R. Ramesh pursued his B.E. degree in Electrical and Electronics Engineering at University of Madras, Chennai, and completed his M.E. degree in Power Systems Engineering at Annamalai University, Chidambaram. He received his Ph.D. degree from Anna University Chennai and has been a faculty of Electrical and Electronics Engineering Department of College of Engineering Guindy, Anna University Chennai since 2003. His areas of interest are Real-time Distributed Embedded Control, On-line power system analysis and Web services.